\documentclass[letterpaper, aps, showpacs, onecolumn]{revtex4}
\usepackage{amsmath}
\usepackage{amssymb}
\usepackage{latexsym}
\usepackage{graphics}
\usepackage{hyperref}

\begin{document}

\title{Quantum Mechanics from the Hamilton-Jacobi Point of View}
\author{Alexander Jurisch}
\affiliation{Physik-Department, Technische Universit\"at M\"unchen, 85747 Garching bei M\"unchen, Germany}
\begin{abstract}
In this article, we develop quantum mechanics upon the framework of the quantum mechanical Hamilton-Jacobi theory. We will show, that the Schr\"odinger point of view and the Hamilton-Jacobi point of view are fully equivalent in their description of physical systems, but differ in their descriptive manner. As a main result of this, a wave function in Hamilton-Jacobi theory can be decomposed into travelling waves in any point in space, not only asymptotically. The well known WKB-theory will be show up as a special case of the more general theory, we develop below. By the example of the linear potential and the harmonic oscillator we will discuss quantum mechanics from the Hamilton-Jacobi point of view. Soft boundary value problems as the connection problem can be solved exactly. Quantizised energies and Maslov-indices can be calculated directely without orthonormalizing wave-functions. Also, we will focus on trajectory-themes, which, in contrast to the Schr\"odinger point of view, follow naturally from the quantum mechanical action function.
\end{abstract}
\maketitle
\section{Introduction}
It must have been a subconscious path, that guided Schr\"odinger to the derivation of his famous equation in the early days of 1926 \cite{Sch}. Schr\"odinger started with the classical Hamilton-Jacobi equation, transformed the classical action function $s$ by $s = -i\hbar\ln[\Psi]$, performed an adventurous variational calculation on the Hamilton-integral and arrived at his celebrated result
\begin{equation}
\Psi''(x) + p^{2}(x)\Psi(x) = 0,\quad p(x) = \sqrt{2m(E-U(x))}/\hbar\quad.
\label{Schrodingerequation}\end{equation}
Several attempts have since been made to relate the quantum mechanics perfectely described by (\ref{Schrodingerequation}) to quantities, that are known from classical mechanics, action and momentum; these attempts are twofold. The first one was taken by Madelung \cite{Mad} already in 1926. Madelung introduced the polar wave-function, $\Psi(x) = A(x)\exp[i S_{{\rm{M}}}(x)/\hbar]$, where $S_{{\rm{M}}}(x)$ is the Madelung action function and both $A, S_{{\rm{M}}}(x)$ are real. Madelungs way was further taken by Bohm \cite{Boh} in his attempt to build a causal quantum theory. But the initial separation of real and imaginary parts of the wave-function leads to very complicated equations for amplitude and phase and will not be pursued here. For exaustive information on the Madelung-de-Broglie-Bohm theory see \cite{Due, Hol1} and for modern application \cite{Wya}. The work of Floyd \cite{Flo} pursued especially the theory of the stationary Hamilton-Jacobi theory with quantum potentials, while Faraggi and Matone \cite{FarMat} probably worked out some fundamental ideas on the geometrical meaning of the quantum potential. In the appendix, we will comment especially on \cite{Flo} in the light of our results.

The second way is to initially keep the real and imaginary parts of the quantum action function together, $\Psi(x) = \exp[i S(x)/\hbar]$, leading to the Hamilton-Jacobi equation of quantum mechanics,
\begin{equation}
(S'(x))^{2} - i\hbar S''(x) = p^{2}(x)\quad.
\label{HJS}\end{equation}
Also in 1926, Wentzel, Kramers and Brillouin \cite{WKB}, starting from (\ref{HJS}), proposed an
expansion of the full quantum action function $S(x,\hbar)$ by small values of $\hbar$, which finally
leads to the classical action, pertubed by small corrections of order $\hbar$. The WKB-theory works
surprisingly well, but it breaks down in the vincity of \emph{quantal} regions, where the WKB-wave function is plagued by
singularities. For an exhaustive review on this topic see \cite{FriTro1} and references therein.

Various methods, most of them are reviewed in \cite{BM}, have been developed to improve the WKB-theory. The most
elaborated among these are the Fr\"oman-Fr\"oman series \cite{Fro} and the use of the quasi-linearization method by
Raghunathan and Vasudevan \cite{RV}. The method of quasi-linearization has the advantage, that it initially does not
rely on the existence of a small expansion parameter, $\hbar$ in our case. Quasi-linearization, as applied to quantum mechanics in \cite{RV} is the starting point of our article. We first develop the relations between the Schr\"odinger theory and the Hamilton-Jacobi theory of quantum mechanics, which, uninterpreted, can already be found in \cite{Sch}. Then, we are going to develop the use of quasi-linearization techniques from a systematic and physically reasonable point of view, which was neglected in \cite{RV}. The systematic use of this technique will lead us to the definition of a quantum correction function. This is a function, that describes all quantum effects of a physical system. Further, the properties of the quantum correction function will be analyzed in detail, especially their relation to the WKB-series expansion. After the properties of the quantum correction functions are clarified, we will apply them to the eigenvalue-problem and the question of quantum trajectories.

In our article, we will develop a quantum theory from the Hamilton-Jacobi point of view. In contrast to the de Broglie-Bohm ansatz, we will not initially separate the amplitude and phase of the wave-function.

\subsection{Structural Notes}
In this subsection, we will examine the interrelation between the Schr\"odinger and Hamilton-Jacobi equation from a
structural point of view. This will clarify, that the solutions of both equations indeed are related to each other by
the well known relations
\begin{equation}
\Psi(x) = \exp\left[\frac{i}{\hbar}S(x)\right],\quad 
S(x) = -i\hbar\ln\left[\Psi(x)\right]\quad.
\end{equation}
By introducing a local gradient field, which can easily be identified with the local quantum momentum,
\begin{equation}
S(x) = \int^{x}P(x)dx \leftrightarrow P(x) = S'(x)\quad,
\end{equation}
the Hamilton-Jacobi equation (\ref{HJS}) can be cast into
\begin{equation}
P^{2}(x) - i\hbar P'(x) = p^{2}(x)\quad.
\label{HJM}\end{equation}

The full solution of the Schr\"odinger equation (\ref{Schrodingerequation}) is a superposition of two linearly independent solutions
\begin{equation}
\Psi(x) = c_{1}\Psi_{1}(x) + c_{2}\Psi_{2}(x)\quad.
\label{Schrodingerwavefunction}\end{equation}
In the language of action, this reads
\begin{equation}
\Psi(x) = c_{1}\exp\left[\frac{i}{\hbar}S_{1}(x)\right] + c_{2}\exp\left[\frac{i}{\hbar}S_{2}(x)\right]\quad.
\label{HJwavefunction}\end{equation}
We want to emphasize here, that (\ref{HJwavefunction}) allows us to decompose a Schr\"odinger wave-function into running waves in \emph{any} point in space, not only asymptotically. The statistical properties of running waves are given by
\begin{equation}
\rho_{i}(x) = \Psi_{i}^{*}(x)\Psi_{i}(x) = \exp[-2\Im{S_{i}(x)}/\hbar],\quad
j_{i}(x) = \pm\Im{\Psi_{i}(x)^{*}\Psi_{i}'(x)} = \pm\Re{P_{i}(x)}\rho_{i}(x)\quad.
\end{equation}

Any of the two linearly independent wave functions solve the Schr\"odinger equation seperately, and any of the two
action functions solve the Hamilton-Jacobi equation seperately. But when we take the logarithm, we find
\begin{eqnarray}
-i\hbar\ln[\Psi(x)] = S(x) &=& -i\hbar\ln\left[c_{1}\exp\left[\frac{i}{\hbar}S_{1}(x)\right] + c_{2}\exp\left[\frac{i}{\hbar}S_{2}(x)\right]\right]\nonumber\\
&\sim& \frac{S_{1}(x)+S_{2}(x)}{2} 
-i\hbar\ln\left[2\cos\left[\frac{S_{1}(x)-S_{2}(x)}{2\hbar}-\frac{\phi}{2}\right]\right]\quad,
\end{eqnarray}
where the $c_{i}$'s have been absorbed in the phase $\phi$.
The function $S(x)$ is again a solution of the Hamilton-Jacobi equation, but $S_{1}(x) + S_{2}(x)$ is not. This fact
is due to the nonlinearity of the Hamilton-Jacobi equation and its physical interpretation is interference.
Usually, the two linearly independent solutions $\Psi_{1}(x), Psi_{2}(x)$ will be complex conjugates, such that $S_{2}(x) =
-S^{*}_{1}(x)$. The superponed wave function, up to an amplitude factor, then reads
\begin{equation}
\Psi(x) \sim \exp\left[-\Im{\{S_{1}(x)\}/\hbar}\right]
\cos\left[\Re{\{S_{1}(x)\}}/\hbar - \frac{\phi}{2}\right]\quad.
\end{equation}

\section{Quantum Correction Functions}
To be able to distinguish between the influence of classical mechanics and quantum mechanics on the behaviour of a
physical system, we propose, that the full quantum momentum field $P(x)$ can be decomposed into its classical part and 
a contribution arising purely by quantum effects. We set
\begin{equation}
P(x) = p(x) + Q(x)\quad.
\label{quasilintransform}\end{equation}
The function $Q(x)$ is called a \emph{quantum} \emph{correction} \emph{function}, because it decribes purely quantum
effects and thus corrects the behaviour of the classical movement, governed by the classical momentum $p(x)$. More
precisely, the function $Q(x)$ describes the difference between the full quantum momentum $P(x)$ and its classical
contribution $p(x)$
\begin{equation}
Q(x) = P(x) - p(x)\quad.
\end{equation}
It can be concluded easily, that $Q(x)$ tends to zero in regions, where the motion is known to be purely classical.\\
To derive an analytic expression for the quantum correction function $Q(x)$, we insert (\ref{quasilintransform}) into the Hamiltion-Jacobi
equation (\ref{HJM}) and find an differential equation for $Q(x)$:
\begin{equation}
Q(x)\{2p(x) + Q(x)\} -i \hbar\{p'(x) + Q'(x)\} = 0\quad.
\end{equation}
This nonlinear first order equation can be formally be integrated to yield a selfconsistent formula for the quantum
correction function $Q(x)$:
\begin{eqnarray}
Q(x) =& -\exp\left[-\frac{i}{\hbar}\int^{x}\{2p(y)+Q(y)\}dy\right]
\int^{x}p'(y)\exp\left[\frac{i}{\hbar}\int^{y}\{2p(z)+Q(z)\}dz\right]dy\quad.
\end{eqnarray}
We emphasize, that this expression for $Q(x)$ is exact; so far, no approximation has been made. In the following, we
will examine the structure of this expression and formulate a systematic perturbational approach. This perturbational
approach does not, as it is usually done for semiclassical questions, start initially with the expansion in terms of
$\hbar$, it will consist of quantum correction functions, that contain every order in $\hbar$. This will open a more
general sight onto the behaviour of physical systems near their semiclassical limit. 

\subsection{Perturbation Theory}
The main reason to examine quantum mechanical systems in Hamilton-Jacobi formulation is to clearly distinguish between classical and quantum behaviour. As we will see below, the semiclassical
limit of Hamilton-Jacobi quantum mechanics is open to perturbation theory, while its full quantum description belongs
to the non-perturbative limit of this theory. The main result of the following subsection is, that in the
one-dimensional case, we are able to introduce an operator formalism, which is, in its structure, 
closely related to the wellknown Dirac- or interaction representation. The central quantity of this operator
formalism is given by  
\begin{equation}
Q(x) = \exp\left[-\frac{i}{\hbar} W(x)\right]{\bf{\hat{Q}}}(x)\exp\left[\frac{i}{\hbar} W(x)\right]\quad,
\label{operatorequation}\end{equation}
where the quantum action function is given by
\begin{equation}
S(x) = s(x) + W(x)\quad,
\end{equation}
$s(x)$ the classical action and $W(x)$ the quantum action correction function.
The structure of (\ref{operatorequation}) enables us to make use of the Baker-Hausdorff identity, which allows us a formal expansion:
\begin{eqnarray}
Q(x) &=& \exp\left[-\frac{i}{\hbar} W(x)\right]{\bf{\hat{Q}}}(x)\exp\left[\frac{i}{\hbar} W(x)\right]\nonumber\\
&=& Q_{0}(x)-\frac{i}{\hbar}\left[W(x),{\bf{\hat{Q}}}(x)\right]
+\frac{1}{2}\left(\frac{i}{\hbar}\right)^{2}\left[W(x),\left[W(x),{\bf{\hat{Q}}}(x)\right]\right] + ...\nonumber\\
&=& Q_{0}(x) + Q_{1}(x) + Q_{2}(x) + ...\quad .
\label{perturbationseries}\end{eqnarray}
The zeroth order correction function $Q_{0}(x)$ comes from
\begin{eqnarray}
Q_{0}(x) &= {\bf{\hat{Q}}}(x)\cdot 1 
= -\exp\left[-\frac{i}{\hbar} 2 s(x)\right]\int^{x}p'(y)\exp\left[\frac{i}{\hbar} 2 s(y)\right]dy\quad.
\end{eqnarray}
This corresponds to the zeroth iteration of equation (\ref{operatorequation}), where the action correction $W(x)$ is chosen zero.\\

\subsection{Convergence Properties}
The convergence behaviour of our perturbation series depends crucially on its key quantity, the function $Q_{0}(x)$. The
whole perturbation series is built on this zeroth order quantum correction function, and thus the behaviour of this
function will influence the behaviour of the perturbation series. The question we have to ask must be, if there are
singularities around classical turning points, which plague the WKB-theory. Anticipating the answer, we can state:  
the function $Q_{0}(x)$ is regular in the vincity of classical turning points $x_{t}$, if, and only if the potential
function $U(x)$ is regular there. The examination now follows.\\
We want to examine, if the integral, which determines the function $Q_{0}(x)$ posseses upper or lower bounds. We have
\begin{equation}
Q_{0}(x)= -\exp\left[i 2 s(x)\right]\int^{x}p'(y)\exp\left[i 2 s(y)\right]dy\quad.
\end{equation}
Since we are interested in the behaviour of $Q_{0}(x)$ in the vincity of a classical turning point, we expand the integrand
around the classical turning point:
\begin{eqnarray}
p'(x) &=& -\frac{U'(x)}{2\sqrt{k^2-U(x)}} \approx -\frac{U'(x_{t})}{2\sqrt{-U'(x_{t})(x-x_{t})}}\quad,\nonumber\\
s(x) &=& \int^{x}\sqrt{k^2-U(x)}dx \approx \frac{2}{3}\sqrt{-U'(x_{t})}(x-x_{t})^{3/2}\quad.
\end{eqnarray}
This approximation is suitable, because the function $p'(x)$, which is the origin of the turning point singularity is
approximately zero almost everywhere, except the turning point region. Thus, the main contribution to the integral comes
directely from the turning point region. Inserting the approximate quantities and performing the integral, we get the
following estimate
\begin{equation}
\lim_{x \rightarrow x_{t}}\left|Q_{0}(x)\right| \approx 
\frac{3^{1/3}}{2^{2/3}}\Gamma\left[\frac{4}{3}\right]\left|\frac{U'(x_{t})}{U^{2/3}(x_{t})}\right|\quad.
\end{equation}
We clearly see, that near the turning point, the behaviour of $Q_{0}(x)$ depends directly on the behaviour of the
potential function $U(x)$. When $U(x)$ is regular at $x_{t}$, which is the case for all potentials, that locally can be
represented by powers $x^{\alpha}$, $\alpha > 0$, the quantum correction function itself is regular and finite. When $U(x)$ is unbounded
near $x_{t}$, which is the case for potentials, that locally can be represented by inverse powers $x^{-\alpha}$, $\alpha >
0$, then the quantum correction function will diverge, too. 

From a more mahematical point of view, Mandelzveig \cite{Man} has shown, that expressions like (\ref{operatorequation}) converge \emph{quadratically}. Physically spoken, quadratical convergence means, that the zeroth iterate $Q_{0}(x)$ is already very close to the exact result $Q(x)$, such that qualitatively the whole quantum mechanical behaviour of a system is to be decribed by the zeroth iterate $Q_{0}(x)$. 

\subsection{The Order of the Pertubing Terms}
From the perturbation series (\ref{perturbationseries}), the order in $\hbar$ is not easily visible. We have seen in (\ref{perturbationseries}), that the full
quantum momentum can be expressed as a series
\begin{equation}
P(x) = p(x) + \sum_{\mu = 0}^{\infty}Q_{\mu}(x)\quad.
\end{equation}
On the other hand, it is known, that the full quantum momentum can be expanded in power of $\hbar$, when $\hbar$ is
small compared to the classical action function:
\begin{equation}
P(x) = p(x) + \sum_{\nu = 1}^{\infty}\hbar^{\nu}P_{\nu}(x)\quad.
\end{equation}
The crucial question is now, how these two series are related to each other. To get more insight into this question,
we start with the selfconsistent expression (\ref{operatorequation}) for $Q(x)$. Due to selfconsistency, the correction function $Q(x)$ is
a function of \emph{all} orders of $\hbar$. To see this more clearly, we formally can put the indices $\mu, \nu$
together and obtain
\begin{eqnarray}
Q(x) &=& \sum_{\mu = 0}^{\infty}Q_{\mu}(x) = \sum_{\mu=0}^{\infty}\sum_{\nu=1}^{\infty}\hbar^{\nu}Q_{\mu
\nu}(x)\quad,\nonumber\\
Q_{\mu}(x) &=& \sum_{\nu=1}^{\infty}\hbar^{\nu}Q_{\mu \nu}(x)\quad.
\end{eqnarray}
This means, that every order of quantum correction functions carries all orders of $\hbar$, as stated above. This
astounding fact will be demonstrated in the next subsection, where the power series in $\hbar$ is calculated
directely from the quantum correction functions. 

\subsection{Derivation of the WKB-Regime}
After these formal preparations, we will calculate the power series in $\hbar$. One of the most interesting results
of this calculation will be a reason for the bad convergence of the WKB-series.\\
The WKB-theory follows from the assumption, that the solutions of the Hamilton-Jacobi equation can be
represented as power series in $\hbar$:
\begin{eqnarray}
S(x) &=& s(x) + \sum_{\nu=1}^{\infty}\hbar^{\nu}S_{\nu}(x)\quad,\nonumber\\
P(x) &=& p(x) + \sum_{\nu=1}^{\infty}\hbar^{\nu}P_{\nu}(x)\quad.
\end{eqnarray}
When these ans\"atze are inserted into the Hamilton-Jacobi equation (\ref{HJS}, \ref{HJM}), there arises an infinite hierarchy of equations,
which can only be solved by successive approximation. Breaking this infinite hierarchy after the first order yields
the familiar WKB result for the wave-function. But the expansion in powers of $\hbar$ is plagued by singularities
around classical turning points, and the strength of these singularities increases by every order in $\hbar$. This is
not the case, when the series expansion is derived from the quantum correction functions, because their expansion in
powers of $\hbar$ will create residual terms of higher order in $\hbar$, which will catch the divergencies and result finite expressions as a
whole.\\ 
Technically, the expansion can be set up as follows. By use of the auxiliary identity $1= p(x)/p(x)$, we get for
the zeroth order
\begin{eqnarray}
Q_{0}(x) &=& -\exp\left[-\frac{i}{\hbar} 2 s(x)\right]
\int^{x}p'(y)\exp\left[\frac{i}{\hbar} 2 s(y)\right]dy\nonumber\\
&=& -\exp\left[-\frac{i}{\hbar} 2 s(x)\right]
\int^{x}p'(y)\left(\frac{p(y)}{p(y)}\right)\exp\left[\frac{i}{\hbar} 2 s(y)\right]dy\quad.
\end{eqnarray}
Performing an integration by parts, one gets
\begin{equation}
Q_{0}(x) = \hbar \frac{i}{2}\frac{p'(x)}{p(x)} + \hbar\frac{i}{2}\exp\left[-\frac{i}{\hbar}2s(x)\right]
\int^{x}dz\left(\frac{p'}{p}\right)'(z)\exp\left[\frac{i}{\hbar}2s(z)\right]\quad.
\end{equation}
Using the auxiliary identity once again and integrating by parts yields
\begin{equation}
Q_{0}(x) = \hbar \frac{i}{2}\frac{p'(x)}{p(x)} - \hbar^{2}\frac{1}{4}\frac{p(x)p''(x)-(p'(x))^{2}}{p^{3}(x)} + R_{0,3}(x)\quad.
\end{equation}
We observe the following facts: the term proportional to $\hbar$ is precisely the first order series expansion
following from the WKB-theory; the following term is of order $\hbar^{2}$, and does not contribute to the first order
in $\hbar$. The residual term, denoted by $R_{\mu=0,\nu=3}(x)$ carries only powers of $\hbar$ equal or larger $\nu =
3$. Such a residual term is not known in WKB-theory, it is lost by initially assuming the power expansion in terms of
$\hbar$.\\
Up to now, we have shown, that the zeroth order quantum correction leads to the correct WKB-term first order in
$\hbar$,
\begin{equation}
\hbar P_{1}(x) = Q_{0,1}(x) = \hbar \frac{i}{2}\frac{p'(x)}{p(x)}\quad.
\end{equation}
The second order expansion in WKB-theory gives
\begin{equation}
\hbar^{2} P_{2}(x) = - \hbar^{2}\frac{1}{4}\frac{p(x)p''(x) - (p'(x))^{2}}{p^{3}(x)} + \hbar^{2}\frac{1}{8}\frac{(p'(x))^{2}}{p^{3}(x)}\quad.
\end{equation}
The first term has already been obtained by the second integration by parts of $Q_{0}(x)$. Further integrations by
parts of $R_{0,3}(x)$ in the same manner would not lead to the required second term, it can only be calculated by
expanding $Q_{1}(x)$ into powers of $\hbar$. The perturbational expression for $Q_{1}(x)$ as calculated above (\ref{perturbationseries}) is
\begin{equation}
Q_{1}(x) = -\frac{i}{\hbar}[\int^{x}Q(z)dz,{\bf{\hat{Q}}}(x)] =
-\frac{i}{\hbar}[\hbar\int^{x}Q_{0,1}(z)dz,{\bf{\hat{Q}}}(x)]\quad.
\end{equation}
The first integration by parts cancels, such that only term of order
$\mathcal{O}(\hbar^{2})$ remain,
\begin{eqnarray}
Q_{1}(x) &=& -\hbar\frac{1}{2}e^{-\frac{i}{\hbar}2s(x)}\int^{x}dz\frac{p'(z)}{p(z)}Q_{0,1}(z)e^{\frac{i}{\hbar}2s(z)} -
[\int^{x}Q_{0,1}(z)dz,{\bf{\hat{R}}}_{0,3}(x)]\nonumber\\
&=& -\hbar^{2}\frac{i}{4}\frac{p'(x)}{p^{2}(x)}Q_{0,1}(x) + R_{1,3}(x) - [\int^{x}Q_{0,1}(z)dz,{\bf{\hat{R}}}_{0,3}(x)]\quad.
\end{eqnarray}
The first order in $\hbar$ is already known from above, $Q_{0,1}(x) = \frac{i}{2}\frac{p'(x)}{p(x)}$, leading to the first term of the desired WKB expression:
\begin{equation}
Q_{1}(x) = Q_{1,2}(x) + ... = \hbar^{2}\frac{1}{8}\frac{(p'(x))^{2}}{p^{3}(x)} + ...\quad.
\end{equation}
With that, we obtain
\begin{eqnarray}
P_{1}(x) &=& Q_{0,1}(x)\quad,\nonumber\\
P_{2}(x) &=& Q_{0,2}(x) + Q_{1,2}(x)\quad,\nonumber\\
..........&&.............................
\end{eqnarray}
The residual terms $R_{\mu\nu}(x)$ und ${\bf{\hat{R}}}_{\mu\nu}(x)$ carry only higher orders in $\hbar$.\\
With this, rather cumbersome, procedure we could show, how all orders of the WKB-series can be obtained, at least in principle. It further clarifies the fidelity of the quantum correction functions: $Q_{0}(x)$ is exact to the first order in $\hbar$, $Q_{0}(x)+Q_{1}(x)$ is exact to second order in $\hbar$ and so on.
Keeping the WKB-form, we can write the quantum mechanical momentum as
\begin{equation}
P(x) = p(x) + \sum_{\nu=1}^{N}\hbar^{\nu}P_{\nu}(x) +
\sum_{\mu=0}^{\infty}\sum_{\nu=N+1}^{\infty}\mathcal{R}_{\mu\nu}(x)\quad,
\end{equation}
where in $\mathcal{R}(x)$ we have subsummized all residual terms. The number $N$ gives the order
in $\hbar$. The residual terms can not be obtained when the WKB-series expansion is set initially. The residual terms result from partially integrating the nonsingular expression for the quantum correction function $Q(x)$, but a nonsingular expression will remain nonsingular. The procedure of partial integration splits the nonsingular quantum correction function into a singular WKB-term and a singular residual term. Both singularities kept together will give a nonsingular function, but omitting the residual terms, as WKB-theory does, will give a badly converging, singular power series in $\hbar$.

\section{The Construction of the Wave-function}
In the introduction section we have seen, that in our theory the wave-function of a system can be written in terms of superponed exponentials, or, if one prefers, in terms of trigonometric functions. Depending on the problem, both representations have advantages and disadvantages. The key to the construction of the wave-function is the action integral, that depends on the direction of integration. In a two turning-point system, we have the choice to integrate from the left ($x_{1}$) to the right ($x_{2}$) or vice versa. We have reserved a whole section to this problem, because its solution is not as easy as it might look at first.

We begin with the wave-function, that is obtained by integrating from the left to the right, $x_{1} \rightarrow x_{2}$. We find, in trigonometric fashion,
\begin{equation}
\Psi(x) = \exp\left[-\int_{x_{1}}^{x}\Im{Q(x')}dx'\right]
\cos\left[\int_{x_{1}}^{x}\left(p(x')+\Re{Q(x')}\right)dx'-\frac{\phi_{\rightarrow}}{2}\right]\quad.
\label{wavefunctionright}\end{equation}
The arrow-index of the reflection phase indicates, that the direction of integration here was chosen as $x_{1} \rightarrow x_{2}$. To construct this wave-function, we have made the following choice for the quantum momentum function
\begin{equation}
P(x) = p(x) + \Re{Q(x)}+i\Im{Q(x)}\quad.
\end{equation}
This choice is of somewhat arbitrary character, we could also have choosen the complex conjugate of $P(x)$ or $-P(x)$ or any other possible combination of signs and complex conjugates. But the choice, once made, gives the system the property of a direction of integration, all other directions of integration have to be consistent with. This becomes clear, when we now construct the wave-function for the direction of integration $x_{1} \leftarrow x_{2}$. The easiest way to do this is to introduce a parity transformation by the operator ${\bf{\hat{T}_{P}}}$, that performs an axialsymmetric transformation. The position of this axis will be the minimum of the potential - the potential itself may be of unsymmetric shape, the minimum is the only condition - that any two turning-point system will possess. We want to know what a wave, that starts at $x_{1}$ would look like, when it starts at $x_{2}$:
\begin{equation}
{\bf{\hat{T}_{P}}}\exp\left[i\int_{x_{1}}^{x}\left(p(x')+\Re{Q(x')}+i\Im{Q(x')}\right)dx'\right]
= \exp\left[i\int_{x_{2}}^{x}\left(p(x')+\Re{Q(x)}-i\Im{Q(x')}\right)dx'\right]\quad.
\end{equation}
The sign of the imaginary momentum function has changed. This is due to the fact, that the real parts of the quantum momentum are even, the imaginary part is an odd function with respect to the minimum of the potential. This can be deduced from (\ref{operatorequation}) by inspection. Thus, when a parity transformation with respect to this minimum is applied, the sign of the imaginary part must be changed. Without this fact, a consistent construction of a wave-function for the direction of integration $x_{1} \leftarrow x_{2}$ is not possible. We finally obtain
\begin{eqnarray}
\Psi(x) &=& \exp\left[\int_{x_{2}}^{x}\Im{Q(x')}dx'\right]
\cos\left[\int_{x_{2}}^{x}\left(p(x')+\Re{Q(x')}\right)dx'-\frac{\phi_{\leftarrow}}{2}\right]\nonumber\\
&=& \exp\left[-\int_{x}^{x_{2}}\Im{Q(x')}dx'\right]
\cos\left[-\int_{x}^{x_{2}}\left(p(x')+\Re{Q(x')}\right)dx'-\frac{\phi_{\leftarrow}}{2}\right]\quad.
\label{wavefunctionleft}\end{eqnarray}

\section{Soft Boundary Problems and Quantization}
In the framework of Hamilton-Jacobi theory, soft boundary problems arise, because the space is separated into classically allowed and classically forbidden regions. By solving the appropriate Hamilton-Jacobi equations, a wave-function $\Psi_{i}$ for each of these regions is known. The boundary value problem is to match all these wave-functions, to obtain a wave-function $\Psi$ defined on the whole space. This boundary value problem is also known as the connection problem, see e. g. \cite{BM}. In the Schr\"odinger point of view the connection problem
\begin{equation}
\Psi_{1}(x) \leftrightarrow \Psi_{2}(x) \leftrightarrow \Psi_{3}(x)
\end{equation}
is a formidable task and will, except some few solvable examples, remain unsovled in general. In the Hamilton-Jacobi point of view, where we can express the wave-functions as exponentials, the connection problem becomes easily treatable on the level of trigonometric and exponential functions:
\begin{equation}
\exp\left[-S_{1}(x)\right] \leftrightarrow 
\exp\left[-\Im{S_{2}(x)}\right]\cos\left[\Re{S_{2}(x)}-\frac{\phi}{2}\right]
\leftrightarrow \exp\left[-S_{3}(x)\right]\quad.
\label{spiritofHJ}\end{equation}
The reflection phase $\phi$ solves the connection problem and with $\phi$ the wave-function can be defined on the whole space.

The points where the wave-function of different spatial regions have to be matched are the classical turning points. The boundary value problem, e.g.
\begin{equation}
\frac{\partial}{\partial x}\ln\left[\Psi_{2}(x)\right]{\bigg{|}}_{x=x_{2}} = 
\frac{\partial}{\partial x}\ln\left[\Psi_{3}(x)\right]{\bigg{|}}_{x=x_{2}}\quad,
\label{boundary1}\end{equation}
is called a soft boundary problem, because it is neither a Dirichlet nor a Neumann boundary problem, but a mixture of both. Writing (\ref{boundary1}) in the spirit of (\ref{spiritofHJ}), we obtain
\begin{equation}
-\Im{Q_{2}(x_{2})}-\Re{Q_{2}(x_{2})}\tan\left[\int_{x_{1}}^{x_{2}} \left(p_{2}(x')+\Re{Q_{2}(x')}\right)dx'-\frac{\phi_{\rightarrow}}{2}\right] = -Q_{3}(x_{2})\quad.
\end{equation}
From this, we obtain the reflection phase
\begin{equation}
\phi_{\rightarrow} = 
2\arctan\left[\frac{\Im{Q_{2}(x_{2})}-Q_{3}(x_{2})}{\Re{Q_{2}(x_{2})}}\right] +
2\int_{x_{1}}^{x_{2}}\left(p_{2}(x')+\Re{Q_{2}(x')}\right)dx'\quad.
\label{phaseLtoR}\end{equation}
For the reverse direction, we find similary
\begin{equation}
\phi_{\leftarrow} = 
2\arctan\left[\frac{\Im{Q_{2}(x_{1})}+Q_{1}(x_{1})}{\Re{Q_{2}(x_{1})}}\right] -
2\int_{x_{1}}^{x_{2}}\left(p_{2}(x')+\Re{Q_{2}(x')}\right)dx'\quad.
\label{phaseRtoL}\end{equation}
The reflection phases, by their dependence on the turning points, are functions of energy.

Knowing the reflection phases, we can easily derive the quantization law by the argument, that both directions of integration must give the same wave-function. The phases of this wave-functions may only differ by $n\pi$. From this, we get
\begin{equation}
\int_{x_{1}}^{x_{2}}\Re{P_{2}(x')}dx' - \frac{\phi_{\rightarrow}+\phi_{\leftarrow}}{2} = n\pi\quad.
\label{quantization1}\end{equation}
Note that by (\ref{phaseLtoR}, \ref{phaseRtoL}) the action-integrals cancel each other in the sum $\phi_{\rightarrow}+\phi_{\leftarrow}$. To make contact with the usual Bohr-Sommerfeld rule, we rewrite (\ref{quantization1}) and introduce the Maslov-index $\mu_{\phi}$
\begin{eqnarray}
\int_{x_{1}}^{x_{2}}p_{2}(x')dx' &=& \pi\left(n + \frac{\mu_{\phi}}{4}\right)\nonumber\quad,\\
\mu_{\phi} &=& \frac{4}{\pi}\left(\frac{\phi_{\rightarrow}+\phi_{\leftarrow}}{2} - 
\int_{x_{1}}^{x_{2}}\Re{Q_{2}(x')}dx'\right)\quad.
\label{quantization2}\end{eqnarray}
With (\ref{quantization2}) the Maslov-index of a physical system is directely calculable by knowledge of the quantum correction function. This is a very valuable result, because quantization in the Hamilton-Jacobi point of view means just to calculate (\ref{quantization2}) to obtain the energy-levels. There is no need to orthonormalize the wave-functions to find the energy-levels, when calculated by (\ref{quantization2}). The theory of Hilbert-spaces ensures, that wave-functions corresponding to eigenvalues of the underlying problem are already orthonormal.

By (\ref{phaseLtoR}, \ref{phaseRtoL}), we can examine another important property of the reflection-phases. They are divided into two parts, one gauge-dependent, one gauge-invariant,
\begin{equation}
\phi = \phi^{{\rm{g}}} + \phi^{{\rm{inv}}},
\end{equation}
Inspection of (\ref{phaseLtoR}, \ref{phaseRtoL}), in combination with (\ref{wavefunctionright}, \ref{wavefunctionleft}) shows, that the lower bound of integration does not necessarily have to be a turning point. Dependent on the direction of integration, which has to be defined at first, the lower bound serves as a point of reference and can be chosen arbitrarily as any point lying between the two turning points. The choice of the lower bound must leave the wave-function invariant. This is ensured by the gauge-dependent part of the reflection-phase, e.g.
\begin{equation}
\phi_{\rightarrow}^{{\rm{g}}} = 2\int_{x_{\rm{ref}}}^{x_{2}}\left(p_{2}(x')+\Re{Q_{2}(x')}\right)dx'\quad.
\end{equation}
The gauge-invariant part is thus given by, e.g.
\begin{equation}
\phi_{\rightarrow}^{{\rm{inv}}} = 2\arctan\left[\frac{\Im{Q_{2}(x_{2})}-Q_{3}(x_{2})}{\Re{Q_{2}(x_{2})}}\right]\quad.
\end{equation}

\section{Quantum Trajectories}
For some reasons, this paper would be the place to set a profound and philosophical discussion on the sense and
usefulness of quantum trajectories into work, but we think that should be done elsewhere. Instead, we want to
demonstrate how easily quantum trajectories are available within the framework of the stationary Hamilton-Jacobi theory. The topic of quantum trajectories is often discussed by introducing a quantum potential and a quantum force. We refuse the introduction of those quantities, because in our opinion they are somewhat misleading. The so called quantum potential is not a potential function as classical potentials are, but depends on the real part of the reduced quantum mechanical action function, so does the quantum force. From this point of view, one anyway has to solve the Hamilton-Jacobi equation at first and calculate the quantum potential afterwards, which makes the introduction of a quantum potential gratuitous (for the here discussed stationary theory). There is another crucial point, that leads us to reject the concept of a quantum potential, because, as we will show below, the classical relation between momentum and velocity does not hold within the framework of quantum mechanical Hamilton-Jacobi theory. The trajectory has thus to be derived from the quantum mechanical action function directely. From this trajectory, by time derivatives, it would in principal be possible to construct a Lagrangian or Newton's equation of motion, but such attempts are, again, gratuitous to our regards.

The use of the stationary Hamilton-Jacobi equation describes isoenergetic dynamics, meaning that motion takes place on a energy-shell in phase-space. The presence of quantum mechanical behaviour has no effect on the energy-shell itself. This can be demonstrated by writing down the the classical and quantum mechanical Hamiltonians
\begin{eqnarray}
\mathcal{H}_{{\rm{cl}}} &=& \frac{p^{2}}{2m} + U(x)\quad,\nonumber\\
\mathcal{H}_{{\rm{qm}}} &=& \frac{P^{2}}{2m} - \frac{i\hbar}{2m}\frac{\partial}{\partial x}P + U(x)\quad,
\label{Hamiltonians}\end{eqnarray}
and comparing them
\begin{equation}
\mathcal{H}_{{\rm{qm}}} - \mathcal{H}_{{\rm{cl}}} = \delta \mathcal{H}\quad.
\label{EnergyConstrained}\end{equation}
If in (\ref{EnergyConstrained}) $\delta\mathcal{H}$ is zero, the energy of quantum and classical motion is constrained to the same energy-shell, quantum and classical motion are isoenergetic to each other. The employment of quasilinearization $P = p + Q$ naturally fullfills 
$\delta\mathcal{H} = 0$ because
\begin{eqnarray}
\mathcal{H}_{{\rm{qm}}} &=& \mathcal{H}_{{\rm{cl}}}\quad,\\
0 &=& 2pQ + Q^{2} - i\hbar\frac{\partial}{\partial x}\left(p + Q\right)\quad.
\label{Isoenergetics}\end{eqnarray}
The presence of the quantum momentum corrections thus have no effect on the energy of the system, they only change the dynamical flow on the energy-shell. For the dynamics that means, that the time the motion between two points takes is changed. In the light of (\ref{Isoenergetics}) it becomes clear how misleading the introduction of concepts like quantum potentials are, because a genuine potential function should of course change the total energy of the system.

The presence of a quantum momentum correction leads to a different trajectory,
\begin{equation}
x_{{\rm{q}}}(t) = x_{{\rm{cl}}}(t) + x_{{\rm{qm}}}(t)\quad,
\label{trajectory}\end{equation}
where the total motion is divised into a purely classical and a purely quantum mechanical part. Naively one would think, that from (\ref{trajectory}) it immedeately follows that
\begin{equation}
m\frac{dx_{{\rm{q}}}(t)}{dt} = p(x) + \Re{Q(x)}\quad,
\label{WrongConnection}\end{equation}
which, as we will show now, is not true. In classical mechanics, the trajectory equation (\ref{trajectoryeq}) and the velocity - momentum relation are equal. We have
\begin{equation}
dt = dx\frac{\partial p(x)}{\partial E} \rightarrow \frac{dx}{dt} = \left(\frac{\partial p(x)}{\partial E}\right)^{-1} = \left(\frac{m}{p(x)}\right)^{-1} = \frac{p(x)}{m}\quad.
\end{equation}
In quantum mechanics, the quantum momentum correction breaks this equality, leading to
\begin{equation}
\frac{dx}{dt} = 
\left(\frac{\partial p(x)}{\partial E} + \frac{\partial \Re{Q(x)}}{\partial E}\right)^{-1} \neq \frac{1}{m}\left(p(x) + \Re{Q(x)}\right)\quad.
\end{equation}
It can easily be seen, that the naive assumption (\ref{WrongConnection}) leads to an obviously wrong result. See \cite{But, Hol2} for statements, that dig in the same hole.\\

The necessary theoretical tool to calculate a trajectory is already known from classical mechanics. As there, we start with the action function
\begin{equation}
\Re{S} = \int_{x_{{\rm{ref}}}}^{x_{2}}dx\left(p(x) + \Re{Q(x)}\right) - E\left(t-t_{0}\right)\quad.
\label{FullAction}\end{equation}
Employing the Jacobi-connection, which is the same as demanding the phase of the wave-function to be stationary, it follows
\begin{equation}
t-t_{0}=\Re{\frac{\partial S}{\partial E}}\quad,
\label{trajectoryeq}\end{equation}
which we will call the trajectory equation. The use of the Jacobi-connection is a natural consequence of the Hamilton-Jacobi theory. Theoretically, it is very well founded and will give the right trajectory for any given action-function. Thus, we dwell on a very safe terrain, whereas (\ref{WrongConnection}) is not more than an assumption.

It is common knowledge, that a standing wave, e.g. a bound state, has a current density of zero, and thus contains
no observable motion. As already stated above, this is true in cases where only the wave-function $\Psi$ is known. The use of
the quantum action function $S$ instead allows a decomposition into running waves, which certainly posses a
dynamical interpretation. This dynamical interpretation allows us to connect the motion of a quantum particle with
a causal picture. To do so, we have to work out the differences between quantum and classical motion at first. From (\ref{trajectoryeq}) we get the running time differences between classical and quantum motion moving from an arbitrarily chosen point to a turning point and back:
\begin{eqnarray}
\Delta t_{\rightarrow} = \int_{x_{{\rm{ref}}}}^{x_{2}}dx\left(\frac{\partial p(x)}{\partial E} - 
\frac{\partial\Re{P(x)}}{\partial E}\right)\nonumber\quad,\\
\Delta t_{\leftarrow} = -\int_{x_{{\rm{ref}}}}^{x_{2}}dx\left(\frac{\partial p(x)}{\partial E} - 
\frac{\partial\Re{P(x)}}{\partial E}\right)\quad.
\label{runningtime}\end{eqnarray}
From (\ref{runningtime}), we obtain the total time-shift between quantum and classical motion
\begin{equation}
\Delta t = \Delta t_{\rightarrow} - \Delta t_{\leftarrow} = 
-2\int_{x_{{\rm{ref}}}}^{x_{2}}dx\frac{\partial\Re{Q(x)}}{\partial E}\quad.
\label{totaltimeshift}\end{equation}
The total time-shift is a positive function of energy and a consequence of the isoenergeticity. Its meaning is the following: the quantum particle moves with a larger local momentum as the classical particle does and reaches the turning point earlier than the classical particle. But for the backward motion the quantum particle is delayed with respect to the classical particle, resulting a positive, total time-shift. The delay of the quantum particle at the turning point stems from the finite quantum momentum there and reflects the finite probability for entering the potential barrier. We emphasize, that a negative sign of the total time-shift (\ref{totaltimeshift}) would result in acausal behaviour, because then the quantum particle would leave before it would have arrived, which is not possible. 
The result (\ref{totaltimeshift}) allows us to derive a remarkable property of quantum trajectories, that is, they leave the phase-space integral invariant. Comparing the classical and the quantum mechanical phase-space integral, this can easily be seen. If we define $T$ as the classical period and $2(T_{1} + T_{2}) = T$, we have
\begin{eqnarray}
\oint p(x) dx - E T &=& \oint \left(p(x)+\Re{Q(x)}\right) dx - 
2 E\left\{\left(T_{1}+\Delta t_{\rightarrow}\right) + \left(T_{2} - \Delta t_{\leftarrow}\right)\right\}\nonumber\rightarrow\\
\oint p(x) dx - E T &=& \oint p(x) dx - E T + \oint \Re{Q(x)} dx + 2 E \Delta t\nonumber\Rightarrow\\
E \Delta t &=& -\frac{1}{2}\oint \Re{Q(x)} dx\quad.
\label{Invariance}\end{eqnarray}
From the last line of equation (\ref{Invariance}) equation (\ref{totaltimeshift}) readily follows. Thus, if isoenergeticity is fulfilled, classical and quantum mechanical motion are described by the same action hypersurface.
With the time-shift (\ref{totaltimeshift}), we now can construct the trajectories
\begin{eqnarray}
\int_{t_{0}}^{t_{1}}dt' &=& \int_{x_{{\rm{ref}}}}^{x}\Re{\frac{\partial P(x')}{\partial E}}dx'\quad(x_{{\rm{ref}}}\rightarrow x_{2})\nonumber\quad,\\
\int_{t_{1}+\Delta t}^{t_{2}}dt' &=& -\int_{x_{{\rm{ref}}}}^{x}\Re{\frac{\partial P(x')}{\partial E}}dx'\quad(x_{{\rm{ref}}}\leftarrow x_{2})\quad.
\end{eqnarray}
By this procedure one obtains quantum trajectories in the classically allowed region, forming a sum of disconnected branches
\begin{equation}
x_{{\rm{q}}}(t) = \sum_{n=0}^{\infty}(-1)^{n}x(t)
\theta\left(t_{n+1} - t_{n} - \Delta t - |t - t_{n+1} + t_{n} + \Delta t|\right)\quad.
\label{subspace}\end{equation}
The trajectories form disconnected branches, because their shape in the classically forbidden region is not available by the trajectory equation (\ref{trajectoryeq}), but below, when the quantum trajectories of the harmonic oscillator will be discussed, we will see how this problem can be solved in a physically reasonable way.

\section{Application}
To demonstrate how quantum mechanics is done from the Hamilton-Jacobi point of view, we will discuss two important systems, that are widely used as models for various physical systems: the linear potential and the harmonic oscillator potential. Especially, we will work out the differences between the Schr\"odinger point of view and the Hamilton-Jacobi point of view. By the ladder, we will obtain an insight into the quantum mechanical behaviour, which can not be gained by the wave-function alone.

\subsection{The Linear Potential}
We consider a linear potential $U(x) = - F x$. By this example, we will focus on the improvement, that is gained by the use of the zeroth order quantum correction function $Q_{0}(x)$ in contrast to the usual WKB-approximation. From the potential, we obtain local the classical momentum and the local classical action as
\begin{eqnarray}
p(x) &=& \sqrt{k^{2} + f x}\quad,\\
s(x) &=&\frac{2}{3 f}\left(k^{2} + f x\right)^{3/2}\quad,\\
k^{2} &=& \frac{2 m E}{\hbar^{2}},\quad f = \frac{2 m F}{\hbar^{2}}\quad.
\end{eqnarray}
The zeroth order quantum corrections function in this case can be calculated analytically, giving
\begin{equation}\begin{aligned}
Q_{0}(x) =& (-i)^{-\frac{1}{3}}\left(\frac{\sqrt{f}}{6}\right)^{\frac{2}{3}}\exp[-i 2 s(x)]\Gamma_{1/3}\left[-i 2 s(x)\right]\quad.
\end{aligned}\end{equation}
During the following, we set $f = 1$ and $k = 0$, because the system is translationally invariant on the energy-axis.
\begin{figure}[h]\centering
\rotatebox{-90.0}{\scalebox{0.4}{\includegraphics{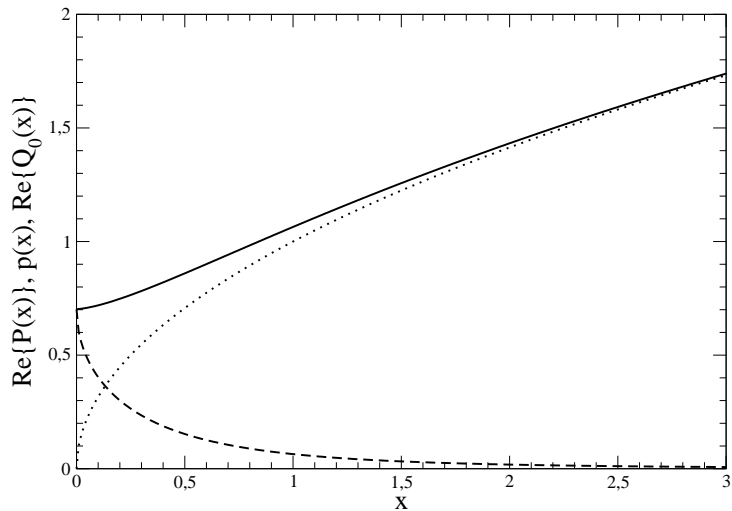}}}
\caption{\footnotesize{Classically allowed phase-space of the linear potential, showing the corrected momentum $\Re{P(x)} = p(x) + \Re{Q_{0}(x)}$ (full line), the quantum correction function $\Re{Q_{0}(x)}$ (dashed line) and the classical momentum $p(x)$ (dotted line). The effect of the zeroth order quantum correction can clearly be distinguished.}}
\label{LinPot1}\end{figure}
In figure (\ref{LinPot1}), we have plotted the phase-space functions of the linear potential. The quantum correction function $\Re{Q_{0}(x)}$, far away from the classical turning point, here $x_{1} = 0$, is negligible and the classical momentum $p(x)$ is in charge. Approaching the turning point region, the quantum correction function $\Re{Q_{0}(x)}$ starts to grow, describing the importance of quantum mechanical behaviour in the vincity of the turning point. At the turning point itself, the real-part of the corrected momentum $\Re{P(x)}$ is finite, while the classical momentum $p(x)$ is zero as usual. This allows only one interpretation: for a classical particle, the potential represents an inpenetrable barrier, and reaching the classical turning point leads undoubtedly to a change of the direction of motion. For a semiclassical particle, described by $\Re{P(x)} = p(x) + \Re{Q_{0}(x)}$, the classical picture is not true anymore. The corrected momentum $\Re{P(x)}$ is finite at the classical turning point, meaning that the potential barrier becomes penetrable with some probability. For a classical particle, the potential barrier represents a hard wall, but for a semiclassical particle, the potential barrier becomes soft. The softness of the potential barrier in the semiclassical case is decribed by the finite momentum functions and by the finite probability to find a particle inside the barrier. We want to emphasize here, that in the WKB-approach the potential barrier remains hard, because of the turning point singularity. The proper connection between classically allowed and classically forbidden region in the WKB-case can only be established by the knowledge of the exact solution of the Schr\"odinger equation, leading to the reflection-phase $\phi = \pi/2$. By the use of quantum correction functions, the turning point singularity is overcome.
\begin{figure}[h]\centering
\rotatebox{-90.0}{\scalebox{0.4}{\includegraphics{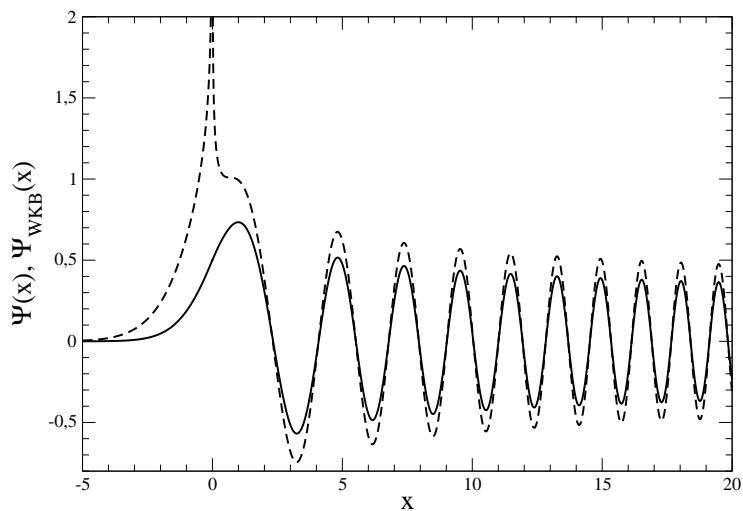}}}
\caption{\footnotesize{Comparison of the wave-function obtained by the zeroth order quantum correction function (full line) and the usual WKB-approximation (dashed line). Already the zeroth order quantum corrections removes the WKB-turning point singularity completely and is, up to the amplitude, identical with the exact solution of the corresponding Schr\"odinger equation, the Airy function.}}
\label{LinPot2}\end{figure}
The linear potential serves as a model for the one turning point case. Taking the turning point  $x_{1}$ as the turning point of the system, the wave-function can be constructed by (\ref{wavefunctionleft})
\begin{equation}
\Psi(x) = \exp\left[-\int_{x}^{\infty}\Im{Q_{0}(x')}dx'\right]
\cos\left[-\int_{x}^{\infty}\left(p(x')+\Re{Q_{0}(x')}\right)dx'-\frac{\phi_{\leftarrow}}{2}\right]\quad.
\end{equation}
The second turning point $x_{2}$ must be set to infinity. This becomes clear when we remember that (\ref{wavefunctionleft}) was derived from the symmetry properties of a two turning point system: the particle comes in from infinity until $x_{1}$ is reached, where it is reflected back to infinity - and at infinity it is reflected again into direction of $x_{1}$. Thus, a one turning point system follows from a two turning point system by setting one of the two turning points to infinity. This procedure is necessary, because it is the only way to take into account the symmetry properties of the quantum correction function as discussed above. Together with (\ref{phaseRtoL}) and the symmetry property of $\Im{Q_{0}(x)}$ we obtain the wave-function
\begin{equation}
\Psi(x) = \exp\left[-\int_{x_{1}}^{x}\Im{Q_{0}(x')}dx'\right]
\cos\left[\int_{x_{1}}^{x}\left(p(x')+\Re{Q_{0}(x')}\right)dx' - \frac{\phi_{\leftarrow}^{{\rm{inv}}}}{2}\right]\quad.
\label{wavefunctionLinPot}\end{equation}
Expression (\ref{wavefunctionLinPot}) perfectely demonstrates how the gauge-dependent part of the reflection-phase cancels, while the gauge-invariant part $\phi^{{\rm{inv}}}$ remains present.
\begin{figure}[h]\centering
\rotatebox{-90.0}{\scalebox{0.4}{\includegraphics{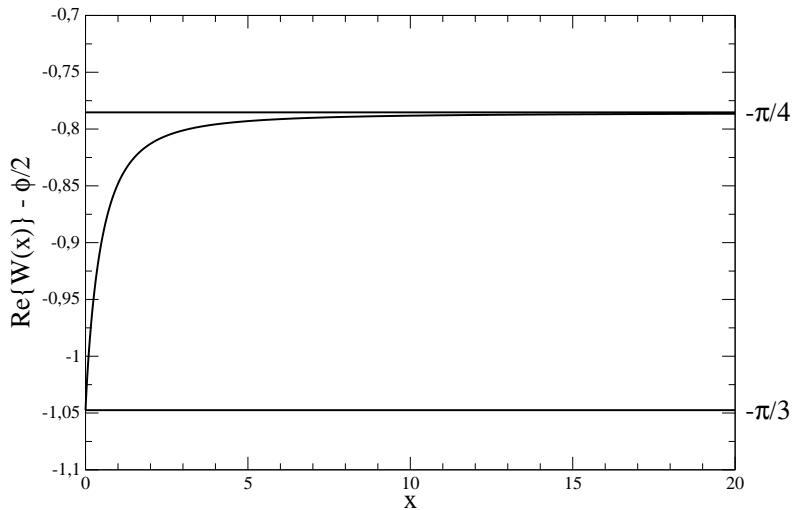}}}
\caption{\footnotesize{Illustration how the correct asymptotic reflection-phase is approached by the interplay of the real part of the quantum action correction and the kinematical phase.}}
\label{LinPot3}\end{figure}
As can be seen in fig. (\ref{LinPot3}) the invariant phase $\phi^{{\rm{inv}}} = 2\pi/3$ is not the WKB reflection-phase $\pi/2$ obtained by the asymptotical expansion of the Airy-function. The asymptotical reflection-phase follows from the limit
\begin{equation}
\lim_{x\rightarrow\infty}\int_{x_{1}}^{x}\Re{Q_{0}(x')}dx' - \frac{\phi_{\leftarrow}^{{\rm{inv}}}}{4} = -\frac{\pi}{4}\quad. 
\end{equation}

\subsection{The Harmonic Oscillator}
The harmonic oscillator is surely the most considered and best understood quantum system. This fact makes it a very interesting test object for the viewpoint on quantum mechanics we haves developed here, because by the harmonic oscillator, we can demonstrate at best the differences between Schr\"odinger and Hamilton-Jacobi quantum mechanics.

We will work in suitably scaled quantities
\begin{equation}
p(x) = \sqrt{k^{2} - \Omega^{2}x^{2}},\quad k=\sqrt{\frac{2mE}{\hbar^{2}}},\quad \Omega = \frac{m\omega}{\hbar}\quad.
\end{equation}
For simplicity, but without loss of generality, we have set $\Omega = 1$ in all numerical calculations.
\begin{figure}[h]\centering
\rotatebox{-90.0}{\scalebox{0.4}{\includegraphics{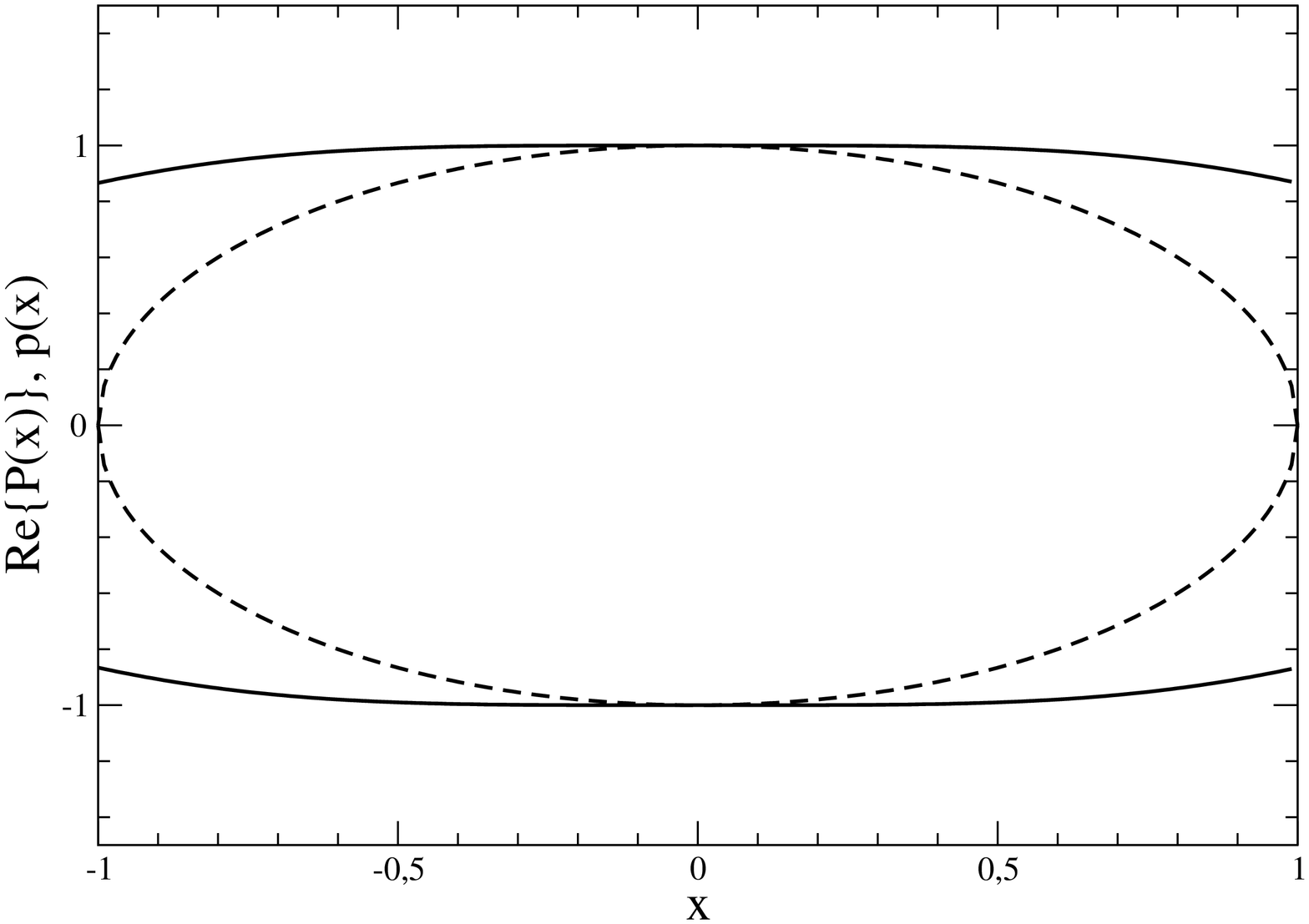}}}
\rotatebox{-90.0}{\scalebox{0.4}{\includegraphics{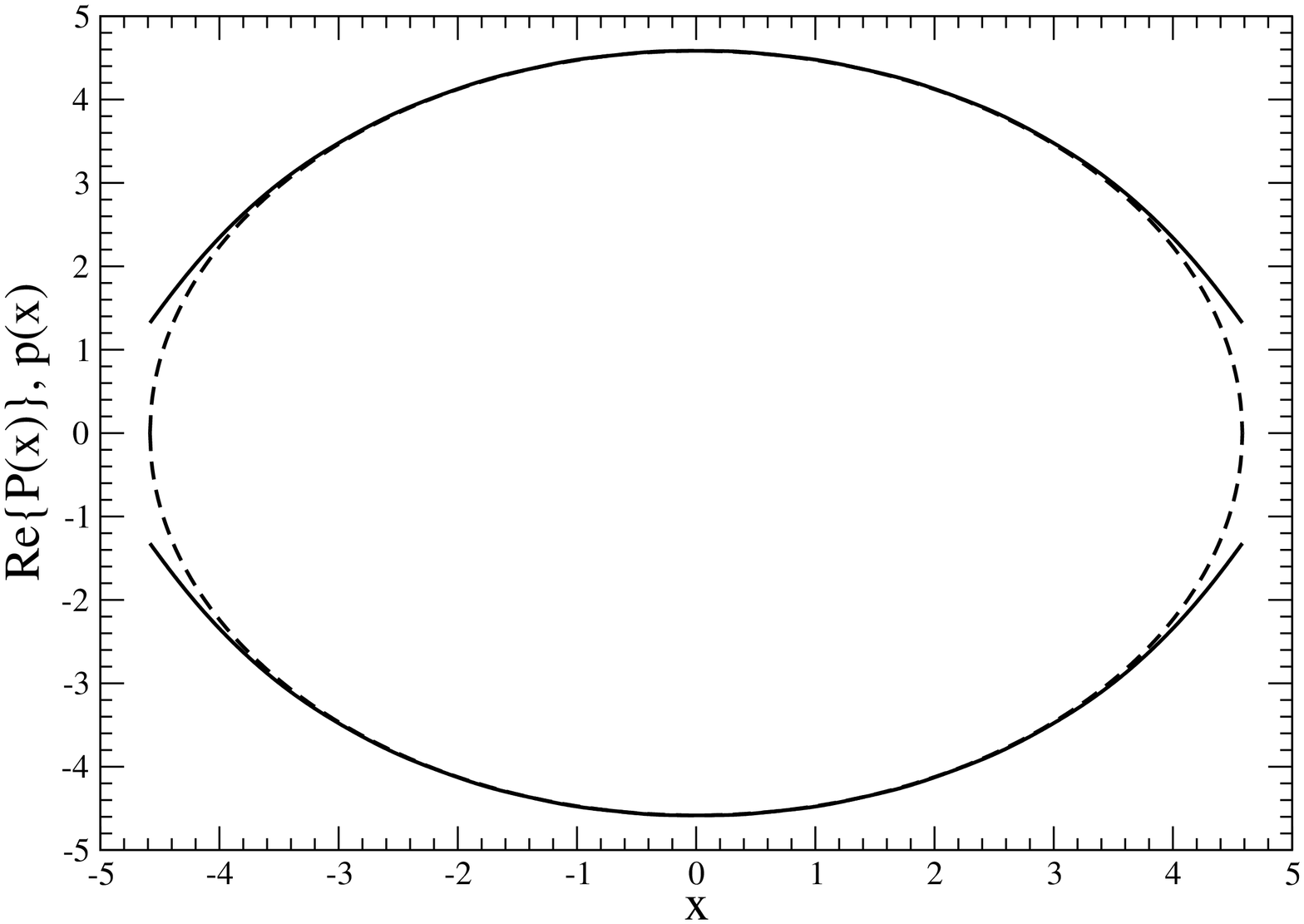}}}
\caption{\footnotesize{Comparison of the phase-space of the harmonic oscillator for the full quantum momentum $\Re{P(x)} = p(x) + \Re{Q(x)}$ (full line) and the classical momentum $p(x)$ for the ground state (left) and the tenth excited state (right). Due to the soft boundary conditions, the quantum mechanical phase-space is an open ellipse, while the classical phase-space is the well-known closed ellipse.}}
\label{HO1}\end{figure}
In figure (\ref{HO1}) we have plotted the quantum mechanical and the classical phase-space of the harmonic oscillator for the ground state energy (left picture). Again we see the typical effect of the quantum correction functions, breaking up the closed classical phase-space, thus making the potential barrier soft and penetrable. The effect of the quantum correction functions is largest for the ground state, as one would naturally expect, because the ground state is close to the anticlassical limit of the system. For the tenth excited state, the right picture in (\ref{HO1}) shows that the influence of $Q(x)$ already has declined, the system has already entered its semiclassical limit and quantum corrections become negligible.

We will not present pictures of oscillator wave-functions here, because we think that up to now the reader might be convinced, that the wave-functions determined by (\ref{wavefunctionright}) or (\ref{wavefunctionleft}) are, up to the normalization amplitude, identical with the oscillator wave-functions determined by the Schr\"odinger equation. Instead we will focus on the calculation of quantities, that are not available within the Schr\"odinger equation. 
\begin{figure}[h]\centering
\rotatebox{-90.0}{\scalebox{0.4}{\includegraphics{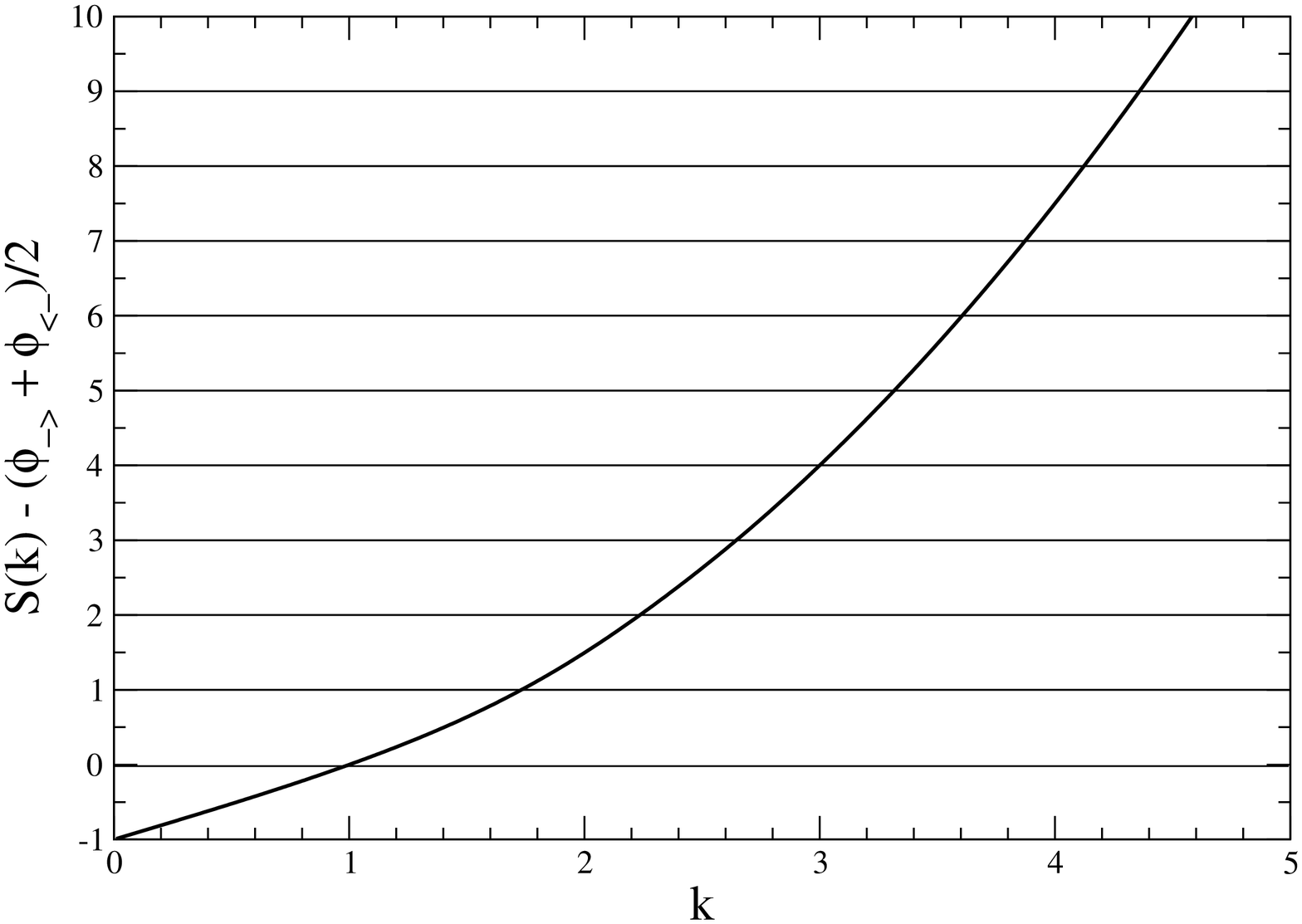}\includegraphics{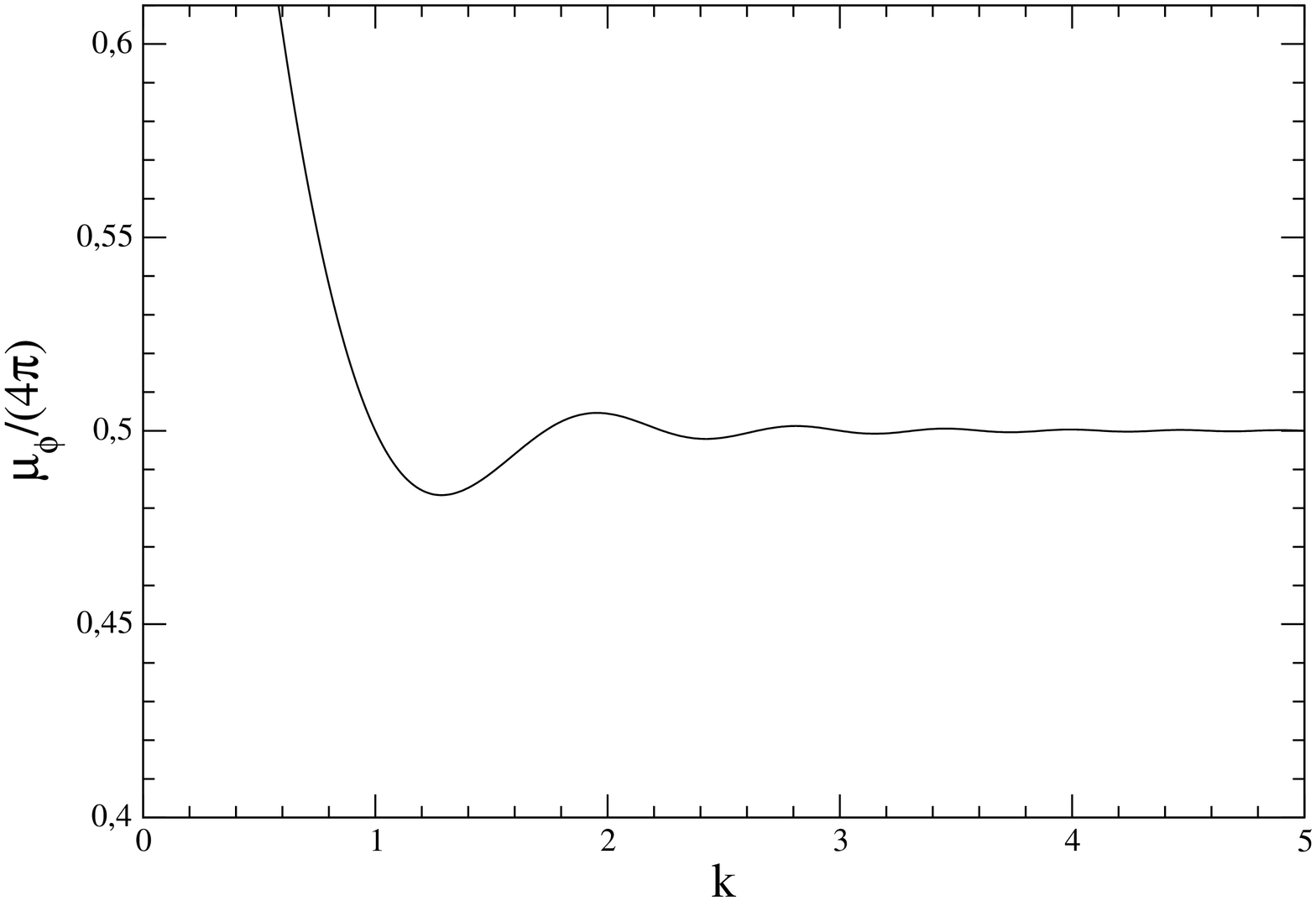}}}
\caption{\footnotesize{Quantization curve and $\mu_{\phi}/4$ for the harmonic oscillator in units of $\pi$.}}
\label{HO2}\end{figure}
In figure (\ref{HO2}) we have calculated the quantization curve and the Maslov-index according the formulae (\ref{quantization1}, \ref{quantization2}). The quantization curve determines the oscillator eigenstates by knowledge of the classical momentum and its quantum corrections. Although the wave-function is implicitly known when the reflection phases (\ref{phaseLtoR}, \ref{phaseRtoL}) are known, there is no need to employ them for determining the spectrum. The knowledge of the quantum correction functions is sufficient for a calculation from first principles. The Maslov-index imposes a tiny oscillating structure onto the quantization curve. The nodes of these oscillations lie perfectely at $1/2$, coincidence with the bound states and recover the Bohr-Sommerfeld rule. This behaviour of the Maslov-index was first calculated in \cite{FriTro2} for the half-sided harmonic oscillator and by a different method.

\subsection{Trajectories for the Harmonic Oscillator}
For the harmonic oscillator, for further details see part I of this paper, we now will discuss the topic of quantum trajectories in detail. We will calculate and discuss trajectories for the ground state and the tenth excited state to illustrate the influence of the quantum corrections functions near the anticlassical and in the semiclassical regime.

\begin{figure}[h]\centering
\rotatebox{-90.0}{\scalebox{0.4}{\includegraphics{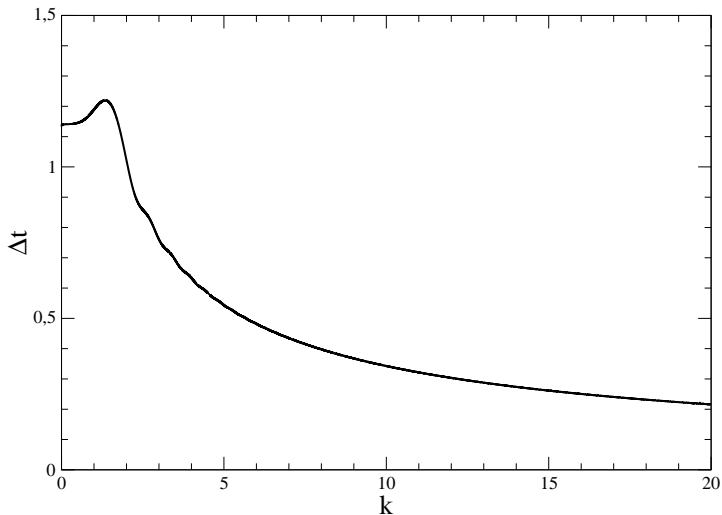}}}
\caption{\footnotesize{Total time-shift $\Delta t$ as function of $k$. For high energies, the total time-shift tends to zero, because the system approaches its classical limit, where quantum and classical trajectories do not differ anymore.}}
\label{HO4}\end{figure}

\begin{figure}[h]\centering
\rotatebox{-90.0}{\scalebox{0.4}{\includegraphics{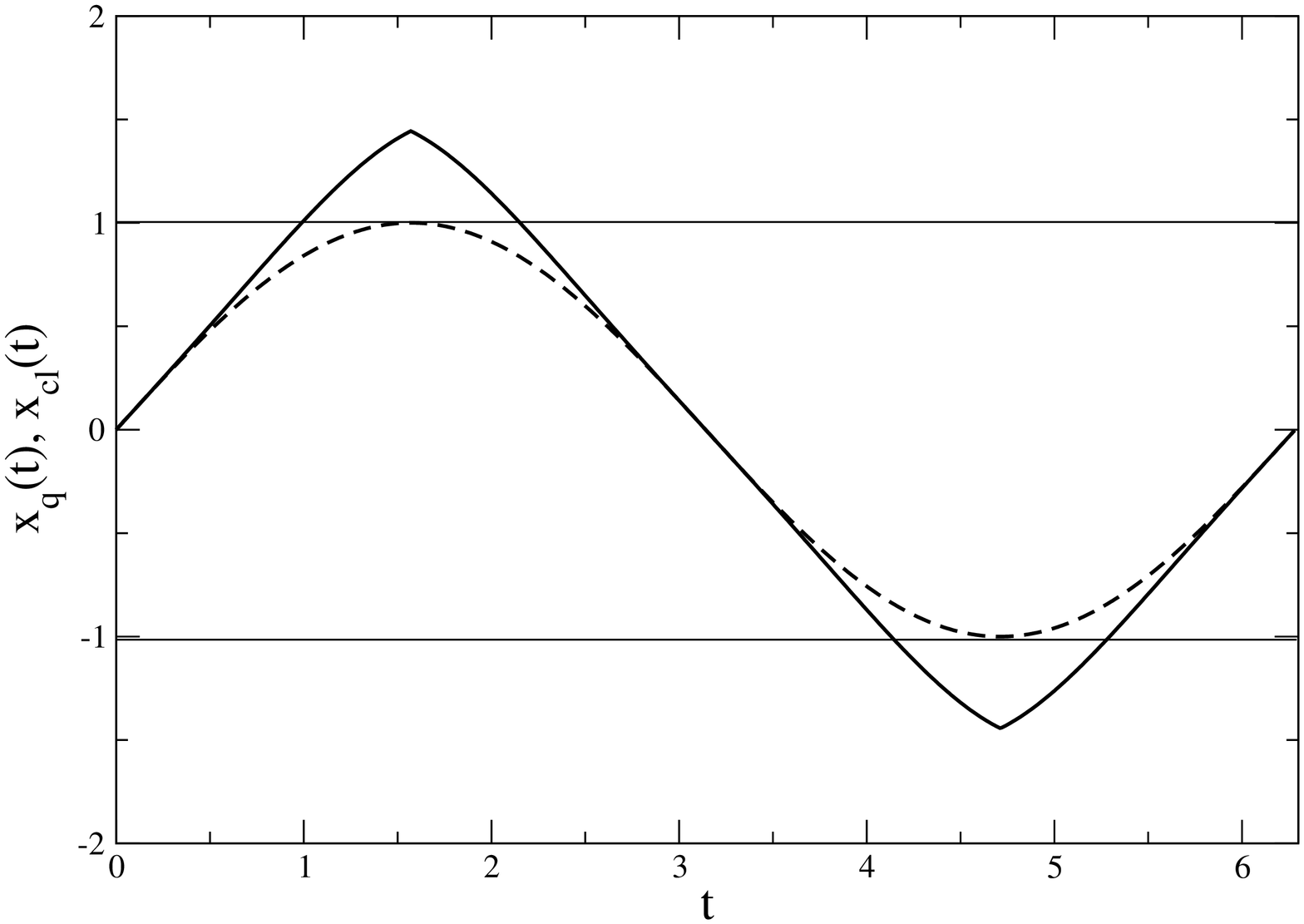}\includegraphics{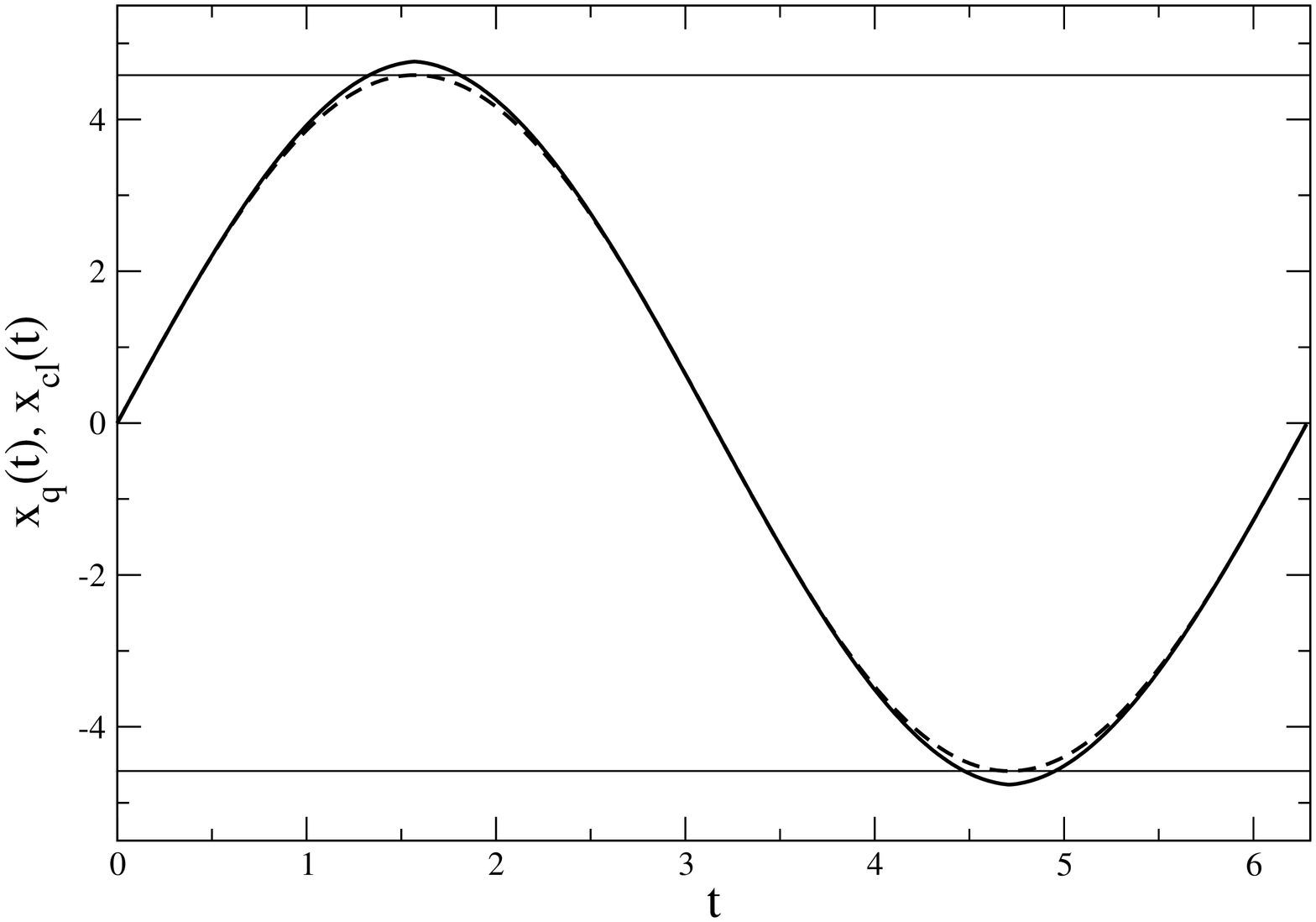}}}
\caption{\footnotesize{Comparison of quantum trajectories (full and dotted lines) and classical trajectories (dashed lines). The full lines are the branches obtained by the trajectory equation (\ref{trajectoryeq}). The total time-shift $\Delta t$ is the amount of time the quantum particle spends in the turning point region. It can be clearly seen, how the limiting process from (\ref{traansatz}) works.}}
\label{HO3}\end{figure}

By (\ref{totaltimeshift}) the time-shift between quantum and classical motion, see figure (\ref{HO4}) is given as
\begin{equation}
\Delta t = -\frac{2}{k}\int_{0}^{x_{2}}\frac{\partial\Re{Q(x)}}{\partial k}dx\quad.
\label{tratimeshift}\end{equation}
The lower bound in (\ref{tratimeshift}) is zero, because we want to calculate the trajectory of a particle starting its journey from the minimum of the oscillator potential $x = 0$. We could have chosen any other point, the choice of the point of reference has no influence on the shape of the quantum trajectory. From the trajectory equation (\ref{trajectoryeq}), we get
\begin{equation}
x_{{\rm{q}}}(t) = \sum_{n=0}^{\infty}(-1)^{n}x(t)
\theta\left(t_{n+1} - t_{n} - \Delta t - |t - t_{n+1} + t_{n} + \Delta t|\right)\quad,
\label{trasubspace}\end{equation}
where the index $n$ denotes the multiple branches obtained by shifting the time by $\Delta t$ in the classically allowed region for the forward and backward motion, respectively. By the trajctory equation (\ref{trajectoryeq}) only the quantum motion between the two turning points can be determined. The cutoff at the turning points is the cause the multiple branches. If we compare (\ref{trasubspace}) with the classical trajectory
\begin{equation}
x_{{\rm{cl}}}(t) = \sqrt{\frac{k^{2}}{\Omega^{2}}}\sin[\Omega t]\quad.
\label{traclassical}\end{equation}
and unse the fact that quantum and classical motion are described by the same frequency, see (\ref{Invariance}), we are lead to make the following ansatz for a quantum trajectory
\begin{equation}
x_{{\rm{q}}}(t) = A_{{\rm{q}}}(k,\Omega, t)\sin[\Omega t],\quad
\lim_{k\rightarrow\infty}A_{{\rm{q}}}(k,\Omega, t) = \sqrt{\frac{k^{2}}{\Omega^{2}}}\quad.
\label{traansatz}\end{equation}
The limit in (\ref{traansatz}) requires, that the quantum amplitude approaches its classical value in the classical limit. The fact that all quantum effects are now located in an amplitude-function reminds us of the path-integral, where the same phenomenon occurs: phase-quantities are given by the classical action function, but the time-evolution amplitude is quantum mechanical. However, by comparison between (\ref{trasubspace}, \ref{traansatz}), we can establish a continous quantum trajectory without cutoffs at classical turning points by analytical continuation beyond the classical turning points, illustrated in figure (\ref{HO3}).
The reader might object this procedure by the argument, that by analytical continuation in this case a kind of information is gained, that does not exist. We agree with this argument to some extent and will therefor call the trajectories in the classically forbidden region apparent trajectories. The present situation is similar to the quantum time-shifts in \cite{JurFri}, where it was found that apparent trajectories enter the classically forbidden region. But the difference between there to present is, that in \cite{JurFri} the potential was a singular one - and no particle, classical or quantum can cross a singularity. But the harmonic oscillator is regular and by the quantum correction functions, the turning point region has become regular, too. Further know we from above, that in the quantum case the potential becomes soft and therefor penetrable. All this put together means the following: for the harmonic oscillator, the reflection probability at a turning point is $|R(k_{n})|^{2} = 1$ for all energies, thus the particle comes back in any case. The particle reaches the classical turning point with finite momentum, which imposes a time-shift in comparison to the classical particle and a phase-shift on the wave-function as a result of the soft boundary condition to be fullfilled. The phase-shift of the wave-function and the time-shift of the motion are two manifestation of the same phenomenon, the relative softness of the potential barrier in the quantum mechanical limit. All these facts allow only one interpretation, the particle as it has arrived at the turning point with finite momentum, goes beyond the turning point and enters the classically forbidden region, appears to be decelerated by the soft barrier, comes to rest and finally moves backward to be expelled again into the classically allowed region. The only explanation thus is the softness-effect of the potential barrier. We emphasize, that this softness has nothing to do with sticking or absorptive reactions, it is a pure quantum mechanical effect. For energies approaching the classical limit, the barrier hardens as is described by the limiting process in (\ref{traansatz}).

\section{Summary and Conclusion}
We have shown how quantum mechanical problems can be solved within the framework of the Hamilton-Jacobi equation of quantum mechanics, where the central quantity of interest is the quantum mechanical action function instead of the wave-function. To our regards, Hamilton-Jacobi theory and Schr\"odinger theory are fully equivalent to each other, they describe the same physics and differ only in their descriptive manner.

In contrast to the well-known Madelung-de-Broglie-Bohm theory, we do not separate real and imaginary part of the action function initially. A very valuable feature of Hamilton-Jacobi theory is, that by use of the quantum action function, the wave-function can uniquely be decomposed into running waves in every point in space, regardless whether the wave-functions describes bound or scattering states. In the Schr\"odinger point of view, this is possible only asymptotically. 

The use of quantum correction functions enopens a deeper insight into the quantum mechanical behaviour of systems, especially in the vincity of turning points, where the local quantum mechanical momentum is finite. This property overcomes the WKB-singulariy at classical turning points and furthermore clearifies the special behaviour of a potential barrier in the quantum mechanical case: the barrier becomes soft and therefor allows the quantum particle to enter it. This softness-property of the potential barrier in the quantum mechanical case is modeled by the soft boundary conditions the wave-function has to obey - and that can be solved exactly by help of the quantum action function - and also explain the finite probability of finding a particle there.

The soft boundary-value problem is solved by a reflection phase. The reflection phase is the key-quantity to obtain the spectrum of a quantum system with help of a quantization curve and allows a direct calculation of Maslov-indices. To obtain the spectrum of a quantum system by this method, only action and momentum functions have to be known, orthonormalized wave-functions are not necessary. Especially for potentials where exact wave-functions are not analytically known, it may be much more simple to calculate the quantization curve instead of searching orthonormalized eigenstates with the Schr\"odinger equation.

Besides the Hamilton-Jacobi theory allows us to calculate quantities that are not available within the Schr\"odinger point of view, the calculation of quantum trajectories is possible. The quantum trajectories obtained by the trajectory equation (\ref{trajectoryeq}) are only defined in the classically allowed region, but by analytical continuation, they can be established also for the classically forbidden region in a physically reasonable manner. The behaviour of the trajectories again mirror the softness-property of the potential barrier in the quantum mechanical case and the finite probability to enter the classically forbidden region is all described by a total time-shift $\Delta t$ (\ref{totaltimeshift}). In contrast to de-Broglie-Bohm theories, no quantum potential is introduced, and therefor no equation of motion. Moreover, the equation of motion in de-Broglie-Bohm theories \cite{Boh, Due, Hol1} and our trajectory equation (\ref{trajectoryeq}) contradict each other. This result was also found by \cite{Flo, FarMat}, but was interpreted differentely. Although the trajectories obtained from the quantum action integral by use of the Hamilton-Jacobi formalism show a somewhat quirky behaviour, they are consistent with our results about the behaviour of the quantum correction function in the vincity of classical turning points and the behaviour of the wave-function. Both, quantum correction functions and trajectories illustrate the softness of a potential function near the anticlassical limit. Moreover, Hamilton-Jacobi formalism is theoretically very well founded and thus we expect ourselves so travel on safe terrain, whereas, as stated in \cite{But, Hol2} the equivalence between velocity and quantum mechanical momentum in de-Broglie-Bohm theories is up to now an unproved assumption.

\begin{appendix}
\section{Comment on Quantum Potential Theories}
The work of Floyd \cite{Flo} interprets stationary Hamilton-Jacobi theory in a very different way as we have done here. Although starting from the same key equation (\ref{HJS}), Floyd developed a completely different theory by separating real and imaginary parts of the quantum action function. By the help of a constrained condition similar to (\ref{Isoenergetics}) he arrived at a nonlinear, second-order equation for the real part of the quantum momentum function
\begin{eqnarray}
E &=& P^{2}(x) + U(x) + {\mathcal{Q}}(x)\quad,\nonumber\\
{\mathcal{Q}}(x) &=& \frac{\hbar^{2}}{2}\left[\frac{P''(x)}{P(x)}-\frac{3}{2}\left(\frac{P'(x)}{P(x)}\right)^{2}\right]
\quad,
\label{FloydHJ}\end{eqnarray}
where ${\mathcal{Q}}(x)$ denotes a quantum potential. If $P(x)$ is the classical momentum $p(x)$, we obtain from (\ref{FloydHJ}) the well-known quantality function, that describes the break-down of WKB-theory in the vincity of classical turning points, see \cite{FriTro1}. But if we are interested in in the quantum momentum, we are not allowed to simply write
\begin{equation}
P(x) = \sqrt{E - U(x) - {\mathcal{Q}}(x)}\quad,
\label{FloydMomentum}\end{equation}
because ${\mathcal{Q}}(x)$ depends on $P(x)$. Moreover, if $P(x)$ has been obtained by (\ref{FloydHJ}), equation (\ref{FloydMomentum}) reduces to an identity $P(x) = P(x)$ which is of no use either. Because of this, we have argued above, that the concept of a quantum potential becomes misleading in stationary Hamilton-Jacobi theory.

Our second objection is, that by transforming (\ref{HJS}) into a real but second order equation, there occurs a loss of information, because for (\ref{FloydHJ}) neither a closed solution formula, nor any initial conditions are known. The apparent arbitrariness for the choice of initial conditions in (\ref{FloydHJ}) lead Floyd to interpret, that there are infinitely many quantum microsates, associated with infinitely many possible initial conditions, that map onto the same energy-eigenstate. For nonlinear, higher order differential equations, its the crux of matter that no unique solution may exist, instead of a large number of different solutions, depending sensitively on initial conditions. But from such a multitude of solutions, the physically relevant one must be extracted. In our first order approach, this is naturally done by (\ref{quasilintransform}, \ref{operatorequation}). Moreover, the action deduced from (\ref{FloydHJ}), the real part of our action function, is assumed to be real on the whole space. Comparing this with our findings above we can state that this is impossible, because a necessary condition for the wave-function to fall off in the classically forbidden region is a purely imaginary action function there. If the concept of the quantum potential would be dropped, this could be deduced from (\ref{FloydHJ}), together with the equations that relate real and imaginary parts of the action function. 

We close this comment and emphasize again, that to our regards, the Hamilton-Jacobi theory of quantum mechanics does not offer a kind of hidden information that is lost in the Schr\"odinger point of view. The information provided by both theories is the same, only their descriptive manner is different.
\end{appendix}

\begin{acknowledgments}
It is a pleasure to thank Harald Friedrich for carefully reading the manuscript and stimulating discussion.
\end{acknowledgments}

\end{document}